\documentclass[preprint,aps,superscriptaddress,nofootinbib,eqsecnum,%
tightenlines,showpacs,showkeys]{revtex4}

\usepackage{epsf}
\usepackage{bm}
\usepackage{amsmath}
\newcommand{\<}{\langle}
\renewcommand{\>}{\rangle}
\renewcommand{\d}{\partial}
\newcommand{\psibar}{\overline\psi}

\newcommand{\bfpi}{\mbox{\boldmath$\pi$}}
\newcommand{\x}{\bm{x}}
\newcommand{\p}{{\bm{p}}}
\newcommand{\q}{{\bm{q}}}
\renewcommand{\P}{\bm{P}}

\newcommand\muai{\mu_{\raisebox{-0.1em}{\scriptsize $I5$}}}
\newcommand\mui{\mu_{\raisebox{-0.1em}{\scriptsize $I$}}}
\newcommand\chir{{\raisebox{0.15em}{$\chi$}}}
\newcommand\sus{{\raisebox{0.15em}{$\chi$}}_{\raisebox{-0.1em}{\scriptsize $I5$}}}
\newcommand\susi{{\raisebox{0.15em}{$\chi$}}_{\raisebox{-0.1em}{\scriptsize $I$}}}
\newcommand\Di{D_{\raisebox{-0.1em}{\scriptsize $I$}}}
\newcommand{\SU}{{\rm SU}}

\begin{document}

\title{Real-time pion propagation in finite-temperature QCD}
%and QCD axial isospin susceptibility}

%\affiliation{Physics Department, Columbia University, New York, 
%New York 10027}
\affiliation{Institute for Nuclear Theory,
University of Washington, Seattle, Washington 98195-1550}
\affiliation{Department of Physics, University of Illinois, Chicago, 
Illinois 60607-7059}
\affiliation{RIKEN-BNL Research Center, Brookhaven National Laboratory,
Upton, New York 11973}

\author{D.~T.~Son}
%\email{son@phys.columbia.edu}
\email{son@phys.washington.edu}
\affiliation{Institute for Nuclear Theory,
University of Washington, Seattle, Washington 98195-1550}
%\affiliation{Physics Department, Columbia University, New York, 
%New York 10027}
%\affiliation{RIKEN-BNL Research Center, Brookhaven National Laboratory,
%Upton, New York 11973}
\author{M.~A.~Stephanov}
\email{misha@uic.edu}
\affiliation{Department of Physics, University of Illinois, Chicago, 
Illinois 60607-7059}
\affiliation{RIKEN-BNL Research Center, Brookhaven National Laboratory,
Upton, New York 11973}
\date{April 2002} 

\begin{abstract}
We argue that in QCD near the chiral limit, at all temperatures below the
chiral phase transition, the dispersion relation of soft pions can be
expressed entirely in terms of three temperature-dependent
quantities: the pion screening
mass, a 
%temperature-dependent 
pion decay constant, and the axial
isospin susceptibility.  The definitions of these quantities
are given in terms of equal-time (static) correlation functions.
Thus, all three
quantities can be determined directly by lattice methods.  
%The axial
%isospin susceptibility is shown to be $f_\pi^2$ at zero temperature
%and nonvanishing at the temperature of the chiral phase transition.
The precise meaning of the Gell-Mann--Oakes--Renner relation at finite
temperature is given.
\end{abstract}
\pacs{%
11.10.Wx, %Finite-temperature field theory
11.30.Rd %Chiral symmetries
}
\keywords{Finite-temperature field theory, chiral symmetries}
%\preprint{CU-TP-1048}
%\preprint{hep-ph/0204226}
\maketitle

\section{Introduction}

Properties of hadrons at high temperatures and densities are of great
interest from both experimental and theoretical perspectives.  One
motivation for studying temperature effects on hadrons comes from the
suggestion that some features of the dilepton spectrum observed in
heavy-ion collisions can be explained by the modification of masses
and widths of mesons by the thermal medium \cite{dilepton}.  
Nevertheless, reliable
information on the temperature modification of hadronic properties is
still lacking.  Lattice simulations, which rely on the imaginary-time
formulation of quantum field theory, have serious difficulties with
real-time quantities.\footnote{Some progress, however, may have 
been made recently \cite{lattice-dilepton}.}
%This formulation is best suitable for the computation of
%equal-time (static) correlation functions, but not of the real-time
%ones.  
The absence of Lorentz invariance at finite temperature implies that
there is no direct relationship between real-time characteristics
of hadrons (for example, the so-called ``pole masses,'' which are
supposedly the positions of poles in propagators) and
quantities that can 
be extracted from Euclidean propagators (e.g., the ``screening
masses,'' which characterize the exponential falloff of static Euclidean
correlators).  Thus, as a rule, lattice measurements of correlation
functions at finite temperature cannot be used to draw conclusions
about real-time propagation of hadrons.

The aim of this paper is to demonstrate that pions present an
exception to this rule.  We shall argue that it is possible to
determine the dispersion relation of soft pions (more precisely, its
real part) at all temperatures below the chiral phase transition, knowing
only equal-time (or static) correlation functions, which, in principle,
can be determined on the lattice.
It should be emphasized that we do
not assume the temperature $T$ to be small compared to the chiral
phase transition temperature $T_c$: we must have $T<T_c$, but $T/T_c$
is allowed to be of order 1.  Our results are valid in the nontrivial
regime where neither perturbative QCD nor chiral perturbation theory
are reliable.  Moreover, our method, with minimal modification,
can be applied to any field theory with a broken symmetry, at
temperatures below symmetry restoration.

That the dispersion relation of a mode can be expressed fully
in terms of static correlation functions is nontrivial, but by no
means unprecedented.  We recall the sound waves, whose
velocity is $u=(\d p/\d\epsilon)^{1/2}$, where $p$ and $\epsilon$ are
the pressure and the energy density, respectively.  The sound speed,
while being a real-time quantity (the pole in the correlator of the
energy density $T^{00}$), can be determined solely from
thermodynamics.  The relation between the speed of sound and the
thermodynamic functions is an exact consequence of the existence of
the hydrodynamic description.  A less familiar example is the
variational Feynman-Bijl formula which relates the phonon spectrum in
superfluid helium to the static density-density correlation function
\cite{BijlFeynman}.  The example most closely related to our problem,
however, is that of spin waves in antiferromagnets \cite{HHPR69}: the
velocity of spin waves at any temperature below phase transition is
equal to the ratio of the stiffness and the magnetic susceptibility in a
direction perpendicular to magnetization.  Both quantities can be
defined from the static response of the system to external fields.
The only difference between QCD and
antiferromagnets is that in the former case the symmetry is
$\SU(2)_V\times \SU(2)_A\simeq {\rm O}(4)$, which is spontaneously broken to
$\SU(2)_V\simeq {\rm O}(3)$, 
while in ferromagnets ${\rm O}(3)$ is broken down to
${\rm O}(2)$ \cite{PisarskiWilczek,RajagopalWilczek}.

%Both
%examples provide essential elements to our understanding of pion
%propagation at finite temperature.  As in the case of phonons,
%hydrodynamics provides the intuition needed for building the effective
%theory of the pions at finite temperature, where the phonons in
%superfluid helium bear resemblance to the pions of QCD as they are
%both Goldstone bosons arising from symmetry breaking.

The paper is constructed as follows.  In Sec.\ \ref{sec:summary} we
summarize the findings of the paper.  In Sec.\ \ref{sec:lagr} we use
simple, but nonrigorous, arguments relying on an effective Lagrangian
to understand these results.  In Sec.\ \ref{sec:hydro} the results are
derived in a more rigorous way from a set of assumptions about the
real-time correlation functions, which comes from hydrodynamics.
In Sec.\ \ref{sec:scalar} we show that our result holds for the
simplest field-theoretical model of a scalar field theory with broken
symmetry.  
%Appendix \ref{sec:sumrule} an attempt to derive 
%the results from
%certain finite-temperature sum rules is presented.  
In the Appendix 
%\ref{sec:z} 
we give a
simple derivation of the known result about the dynamical critical exponent
$z$, and derive the critical scaling of a diffusion coefficient.

\section{Summary of results}
\label{sec:summary}

We claim that, in QCD with two light flavors, at temperatures below
the chiral phase transition, the real part of the dispersion relation
of sufficiently soft pions is given by the following equation:
\begin{equation}
  \omega_\p^2 = u^2 (\p^2 + m^2) \,. 
%\qquad p \equiv |\p| \,.
  \label{omega}
\end{equation}
In this paper the following terminology is used: $u$ is the pion {\em
velocity} (although, strictly speaking, it is the pion velocity only
when $m=0$), and $m$ is the pion {\em screening mass} (we shall show
that it is the same screening mass as defined on the lattice).  The
energy of a pion at $\p=\bm{0}$, $m_p=um$ is called the pion {\em pole
mass}.

At finite temperature, the meaning of soft pions may need some
clarification.  We shall understand Eq.\ (\ref{omega}) as a statement
that the correlators of operators carrying pion quantum numbers have a
pole at the frequency with the real part determined by Eq.\
(\ref{omega}).

The pion velocity $u$ is the ratio of two statically measurable
quantities, the {\em temperature-dependent} pion {\em decay constant} $f^2$
and the {\em axial isospin susceptibility} $\sus$,
\begin{equation}
  u^2 = {f^2\over\sus} \, .
  \label{u}
\end{equation}
The axial isospin susceptibility $\sus$ can be defined as the second
derivative of the pressure with respect to
 the axial isospin chemical potential
(see Sec.\ \ref{sec:lagr}).  Equivalently, it can be defined 
via the
static {\em Euclidean} correlator of the axial isospin charge densities,
\begin{equation}
  \delta^{ab}\sus   = \int\limits_0^{1/T}\!d\tau\int\!d\x\, 
  \< A_0^a(\tau,\x)A_0^b(0,{\bf 0})\>\,,
  \label{chi}
\end{equation}
where
\begin{equation}
  A_0^a \equiv \psibar\gamma^0\gamma^5{\tau^a\over2}\psi
  \label{chi1}
  \,,
\end{equation}
$\psi$ is the quark field, $a,b=1,2,3$, $\tau^a$ are 
isospin
Pauli matrices, ${\rm Tr}\,\tau^a\tau^b=2\delta^{ab}$, and
$\<\cdots\>$ denotes thermal averaging, which can be taken by evaluating
a Euclidean Feynman path integral with appropriate boundary conditions.
The quantity $\sus$
has been considered previously \cite{Hansen}.

The pion decay constant $f$ and the screening mass $m$ can be
determined from the static Euclidean pion correlators at 
small momenta, 
which are predicted to have the form
\begin{equation}
  \int\limits_0^{1/T}\!\!d\tau\int\!d\x\, e^{-i\q\cdot\x} 
  \<\varphi^a(\tau,\x)\varphi^b(0,{\bf 0})\>=
  {1\over f^2} {\delta^{ab}\over \q^2+m^2} \, ,
  \label{fm}
\end{equation}
where the scalar field $\varphi^a$ is defined as
\begin{equation}
  \varphi^a \equiv {i\psibar\gamma^5\tau^a\psi\over
  \<\psibar\psi\>} \,.
  \label{phipsi}
\end{equation}
Equation (\ref{fm}) is supposed to be valid when $|\q|\ll m_\sigma$, 
where $m_\sigma$ is the (temperature-dependent) screening mass of 
the $\sigma$ meson.

The way $m$ enters Eq.\ (\ref{fm}) explains why we term it the pion
screening mass.  As far as $f$ is concerned, 
the conventional definition of the pion decay constant,
$f_\pi p_\mu=\<0|A_\mu|\pi(p)\>$, cannot be used at finite
temperature, since neither the vacuum state $|0\>$ nor the one-pion
state $|\pi\>$ allows generalization in thermal media. 
Both $\sqrt{\sus}$ and $f$ approach $f_\pi$ as $T\to0$, so both can be
viewed as the generalization of $f_\pi$ to finite temperature.  We,
however, reserve the name ``pion decay constant'' for the quantity 
$f$ defined in Eq.\ (\ref{fm}).\footnote{See Sec.\
\ref{sec:hydro-comp} for a discussion related to this point.}
In contrast to the susceptibility 
$\sus$, $f$ cannot be defined at temperatures above critical.

We shall also show that the Gell-Mann--Oakes--Renner (GOR) relation
can be generalized to finite temperature, and does in fact become two separate
relations for the screening and pole masses of the pions,
\begin{equation}
  f^2 m^2 = \sus m_p^2 = - m_q \<\psibar\psi\> \,.
  \label{GOR}
\end{equation}
For simplicity, in this paper we assume $m_u=m_d=m_q$.

Equation (\ref{u}) is the direct analogue of a similar equation for the
velocity of spin waves in a quantum antiferromagnet at temperatures
below the phase transition \cite{HHPR69}.  The counterpart of $f$ is the
stiffness (denoted as $\rho_s$ in Ref.\ \cite{HHPR69}), while
$\sus$ is similar to the magnetic susceptibility in a direction 
perpendicular to magnetization.

An important point we wish to emphasize is that the relations 
(\ref{omega})--(\ref{GOR})
are exact in the chiral limit, i.e., the limit when $m_\pi$ (vacuum
pion mass) and $\p$
are infinitesimally small, at any temperature in the interval from
0 to $T_c$. Although the methods we use to derive them may look
similar to chiral perturbation theory, unlike the chiral
perturbation theory, we do not consider perturbations around
$m_\pi=0$, $T=0$, treating both $m_\pi$ and $T$ as small parameters.
Rather, we consider perturbations around $m_\pi=0$ at a fixed
$T$, the latter not assumed to be small. Therefore, unlike in chiral
perturbation theory, we are not able to calculate the temperature
dependence of parameters such as $\chi$, $f$, $u$, or $\psibar\psi$.
However, we are able to show that their temperature 
dependence must be such that relations (\ref{omega})--(\ref{GOR}) hold.

Given these equations, exact in the chiral limit, it is legitimate to
ask how useful these relations can be at the physical value of
the pion mass $m_\pi=140$ MeV. To answer this question quantitatively,
a (perhaps lattice) calculation of the $T$ and $m_\pi$ dependence is needed, 
which is beyond the scope of this paper.
However, we shall attempt to give a semiquantitative answer to this
question.
% in the concluding Section of the paper. 

It is easier to begin with $T=0$. In this case formula 
(\ref{omega}) becomes trivial: the dispersion relation
is exact, simply by Lorentz invariance, for any $\p$ and $m=m_\pi$.
It is a straightforward exercise to show, using PCAC
(partial conservation of axial vector current), that
to leading order, i.e., ${\cal O}(m_\pi^2)$, the correlator in Eq.\ (\ref{chi})
whose value at $m_\pi=0$ is $\sus$, does not depend on $m_\pi$. Thus
one can expect that measuring the correlator (\ref{chi}) even
at the physical $m_\pi$ (or higher, as is typical in a lattice
calculation due to the high price of simulating with light quark masses) 
is not significantly different from $\sus$. In order to extract
the value of $f$ from the relation (\ref{fm}) reliably, one requires
that higher mass states in the channel with pion quantum numbers
do not significantly contaminate the exponential falloff
of the correlator in coordinate space. Since the masses of such states
are typically above 1 GeV, the exponential tail contribution from the 
light state of mass $m_\pi=140$ MeV should be easy to separate.

At nonzero $T$ the region of applicability of Eq.\ (\ref{u}) is limited
by terms of higher powers in $\p$. For simplicity, in the exact chiral
limit $m_\pi=0$ the dispersion relation takes the form
$\omega=|\p| - {i\over2}D'\p^2 + \cdots$, where $D'$ is
a temperature-dependent parameter (see Sec.~\ref{sec:hydro} and the Appendix). 
The condition on momenta
which is required to neglect nonlinearity in the dispersion relation
is $|\p|\ll 1/D'$. Calculation of the diffusion constant $D'$
is a challenging task. Taking an estimate from Ref.~\cite{Smilga},
$D'= C T^3/f_\pi^4$, with a numerically rather
small $C\sim0.1$, we can conclude
that even for $T\sim f_\pi$ at momenta of order $|\p|\sim 100$ MeV
the nonlinearity in the dispersion relation is still small.

As $T=T_c$ the dispersion relation is essentially nonlinear, which
manifests itself in the divergence of $D'$ as $T\to T_c$ (see the
Appendix). As $T\to T_c$ the maximum momentum at which the dispersion
relation can be considered linear (in the chiral limit) decreases and
vanishes at $T_c$. Although the power with which the width of the
linearity window shrinks to zero can be determined (by extension
of arguments given in the Appendix), the
pre-exponent is unknown, and a nonperturbative calculation is, in
principle, required to determine quantitatively the size of the
nonlinearity in the dispersion relation at a given $T$ and
$|\p|$. Such a calculation is beyond the scope of the paper.

Equations (\ref{omega})--(\ref{GOR}) were used in Ref.\
\cite{critical_pions} to extract information about the critical
behavior of the pion velocity and masses near the critical
temperature.  A brief derivation of Eqs.\
(\ref{omega})--(\ref{GOR}) was also sketched in Ref.\
\cite{critical_pions}.  We present a more extended version of this 
derivation in Sec.~\ref{sec:lagr}.

In Sec.~\ref{sec:hydro} we provide a new derivation of the
relations (\ref{omega})--(\ref{GOR}) 
using the operator approach based on hydrodynamic
equations. For this purpose we shall need expressions for $\sus$
and $f$ in terms of equal-time rather than static (zero frequency)
correlators. These relations follow from definitions (\ref{chi})
and (\ref{fm}) 
in the exact chiral limit ($m_\pi=0$) or when the temperature is high enough 
($T\gg m_\pi$):
%In the exact chiral limit ($m_\pi=0$), or when the temperature is high enough 
%($T\gg m_\pi$), the axial isospin susceptibility can also be expressed
%via the {\em equal-time} correlator of $A^a_0$,
\begin{equation}
  \delta^{ab}\sus = {1\over T}
  \int\!d\x\, \< A_0^a(t,\x) A_0^b(t,{\bf 0}) \>\,.
  \label{chi-eq}
\end{equation}
In the chiral limit, the equality of the 
static correlator [Eq.\ (\ref{chi})] and the equal-time
correlator [Eq.\ (\ref{chi-eq})]
of $A_0^a$ is a
consequence of the conservation of the axial isospin charge.
Similarly,
%In the exact chiral limit of $m_\pi=0$ or at high temperatures,
%$T\gg m_\pi,|\q|$, one can rewrite the left hand side of Eq.\ (\ref{fm}) as
%an equal-time correlator,
\begin{equation}
  {1\over T} \int\!d\x\, e^{-i\q\cdot\x} 
  \<\varphi^a(t,\x) \varphi^b(t, {\bf 0})\> =
  {1\over f^2} {\delta^{ab}\over \q^2+m^2} \,.
  \label{fm-eq}
\end{equation}

Note that, in contrast to Eqs.~(\ref{chi}) and (\ref{fm}), 
Eqs.~(\ref{chi-eq}) and (\ref{fm-eq}) require an additional condition
$T\gg m_\pi$. This condition is, however, not
so dramatic, if one recalls that the relavant energy scale
of order $T$ is the lowest nonzero Matsubara frequency, $2\pi T$. 
Thus, in practice,
one requires $2\pi T\gg m_\pi$, which is satisfied reasonably well
for $T$ of order $T_c\approx 160$ MeV and the physical pion mass. 

However, a much 
stronger condition is required for
the  applicability of the hydrodynamic description:
$|\p|$ (and $m_\pi$) must be smaller than the
typical collision rate $\tau^{-1}\sim T^5/f_\pi^4$ 
(according to Ref.~\cite{Goity}). At small $T\ll f_\pi$ this condition
is much stronger than the conditions that are required for the
effective Lagrangian derivation to hold (typically $|\p|,m_\pi\ll
m_\sigma$ --- see Sec.~\ref{sec:lagr}). 
However, both derivations apply in the
required regime infinitesimally close to the chiral limit.
The fact that the hydrodynamic derivation has a smaller
validity range, however, does not mean that the result is not
valid outside this range. As an example, consider the
case of $T=0$, where the results (\ref{omega}), (\ref{u}) hold trivially,
while hydrodynamics does not apply at all. A less trivial
example is presented in Sec.~\ref{sec:scalar}, where explicit
calculation verifies Eq.\ (\ref{u}). This weak-coupling calculation,
however, does not require any notion of hydrodynamics.

The advantage of the hydrodynamic approach is that it allows us
to consider properly the effects of the dissipation, which 
are inherently beyond the Lagrangian approach.
Before proceeding to the new operator derivation (Sec.\
\ref{sec:hydro}) we review the derivation based on the effective
Lagrangian approach.

\section{Effective Lagrangian approach}
\label{sec:lagr}

\subsection{Axial isospin susceptibility and $f_\pi$ at $T=0$}
\label{sec:mui5}

Since an important role in our analysis is played by the axial isospin
susceptibility, we first consider this quantity at zero temperature
and its relation to the pion decay constant.
The quark part of the QCD Lagrangian at finite axial isospin chemical
potential $\muai$ is given by
\begin{equation}
  {\cal L}_{\rm quark} = i\psibar \gamma^\mu D_\mu \psi 
  - m_q \psibar\psi
  + \muai \bar\psi\gamma_0\gamma_5{\tau_3\over2}\psi \,,
\label{lquark}
\end{equation}
where $D_\mu$ is the color covariant derivative.  This chemical
potential $\muai$ is coupled to the axial isospin charge $A^3_0$
[cf.\ Eq.\ (\ref{chi1})], which
generates the $\SU(2)_A$ part of the 
$\SU(2)_V\times \SU(2)_A$ chiral
symmetry.

The response of the QCD vacuum to $\muai$ can be found 
%in the way analogous to the isospin
%chemical potential, $\mui$ \cite{mui}, 
from the effective chiral Lagrangian.
%We shall mainly be interested
%in the corresponding susceptibility, i.e.,
%the second derivative of the vacuum energy with respect to
%$\muai$.  
%However, the technique allows one to determine
%the effect of $\muai$ on masses of low energy excitations,
%(pions in this case) as well. 
The latter, to lowest order of momenta, masses, and chemical potential,
is completely fixed by the chiral symmetries and is given by
\begin{equation}
  {\cal L}_{\rm eff}= 
  {f_\pi^2\over4} {\rm Tr}\, \nabla^\nu \Sigma\nabla_\nu \Sigma^\dagger
%  - m_\pi\<\bar\psi\psi\>_{\rm vac} 
  + {f_\pi^2 m_\pi^2\over2}
  {\rm Re }\, {\rm Tr}\, \Sigma \, ,
\label{leff}
\end{equation}
where $\Sigma$ is an $\SU(2)$ matrix whose phases
% according to $\Sigma\to U_L\Sigma U_R^\dagger$
describe the pions, $\Sigma = e^{i\tau^a\pi^a/f_\pi}$,
and $\nabla$ denotes the covariant derivative, which is defined as
\begin{equation}
  \nabla_0 \Sigma = \partial_0 \Sigma - {i\over2}\muai
  (\tau_3\Sigma + \Sigma\tau_3)\,, 
\quad
%\mbox{and}\quad
\nabla_i \Sigma = \partial_i \Sigma\,,
\quad
%\mbox{for 
i=1,2,3 \, .
\label{deriv}
%  ( I_3\Sigma - \Sigma I_3),
\end{equation}
% where $I_3$ is the generator of the third component of isospin of a
% quark field $I_3 = {\rm diag}(1/2,-1/2)$.
% \begin{equation}
% I_3 = \mat +1/2&0\\0&-1/2\emat.
% \end{equation}
At $\muai=0$, the Lagrangian (\ref{leff}) is the standard chiral
Lagrangian with two phenomenologically determined constants $f_\pi$
and $m_\pi$.  The way $\muai$ enters the effective description is
completely fixed by symmetries to lowest order.  This can be seen by
promoting the $\SU(2)_A$ symmetry to a local symmetry and treating
$\muai$ as the time component of the $\SU(2)_A$ vector potential
\cite{mu2}.

The Lagrangian (\ref{leff}) and its derivation is analogous to the
case of the
effective Lagrangian at finite (vector) isospin chemical potential $\mui$
considered in Ref.\ \cite{mui}.  
A significant difference between the cases studied here and in
Ref.\ \cite{mui} is that
the QCD {\em vacuum} breaks the $\SU(2)_A$ (axial isospin)
symmetry {\em spontaneously}, but
remains symmetric under the $\SU(2)_V$ (vector isospin) symmetry.
%At the level of the
%effective Lagrangian, the vacuum state, $\bar\Sigma=1$, is not
%invariant under the $SU(2)_A$ rotations: $U_L \bar\Sigma U_R^\dagger \ne
%\bar\Sigma$ when $U_L=U_R^\dagger$. 
It is important to note, however, that the $\SU(2)_A$ {\em is} a
symmetry of the Lagrangian (at $m_q=0$), as good as the
$\SU(2)_V$.  In particular, the axial isospin
current $A^a_\mu$ is conserved in the chiral limit.  Thus, it is entirely
legitimate to consider the theory at finite $\muai$ and use
symmetry arguments to fix the $\muai$ dependence of the effective
Lagrangian.
%This allows us to fix uniquely the dependence on
%in Eq.\ (\ref{leff}).

The vacuum energy density depends nontrivially on $\muai$ already for
arbitrarily small $\muai$. (This is in contrast to the case of the
isospin chemical potential $\mui$, where $\mui$ needs to be larger
than a threshold equal to $m_\pi$ in order to change the ground
state.)  The isospin axial susceptibility, at $\muai=0$, is easy to
determine using the effective Lagrangian (\ref{leff}):
\begin{equation}\label{chifpi}
\sus\equiv{\partial^2 {\cal E}_{\rm vac}\over\partial \muai^2}
 \Biggl|_{\muai=0}
= - \left.
{\partial^2 {\cal L}_{\rm eff}\over\partial \muai^2}\right|_{\Sigma=1}
= f_\pi^2\,.
\end{equation}

This result is of potential importance for lattice QCD calculations
because, in principle, it allows one to determine $f_\pi$ directly by
measuring the axial isospin susceptibility.  To our knowledge, this
has not been done on the lattice for temperatures below the chiral
phase transition.  The isospin susceptibility (as well as the
isoscalar, i.e., baryon number susceptibility) measurement has been
done using staggered fermions in \cite{gottlieb}. In the formulation
used in Ref.\ \cite{gottlieb}, introducing the axial isospin chemical
potential would require replacing $\exp(-\mu a)$ on the links (where
$a$ is the lattice spacing) with $\exp(-\mu a \zeta_x)$, where
$\zeta_x$ is the usual staggered factor $ \zeta_x\equiv
(-1)^{x_1+x_2+x_3+x_4}$ --- the representation of $\gamma_5$ in
the staggered fermion action.

\subsection{Pion velocity}

To obtain Eq.\ (\ref{u}), we expand the previous discussion to nonzero
temperature.  We first presume that the dynamics of the pions is
described by some
%(Euclidean) 
effective Lagrangian ${\cal L}_{\rm eff}$.  Strictly speaking, this is
not correct since dissipative effects cannot be included in the
effective Lagrangian.  We can expect to recover the 
correct answers if,
in the infrared, the pion thermal width is
negligible compared to its energy.  This has been seen
in explicit calculations at low $T$, where chiral perturbation theory
can be used \cite{Shuryak,Goity}.  We furthermore
assume that this Lagrangian is local and can be expanded in powers of
momenta. To lowest order, the Lagrangian is fixed by symmetries up to
three coefficients, $f_t$, $f_s$, and $m$,
\begin{equation}
  {\cal L}_{\rm eff}= 
  {f_t^2\over4} {\rm Tr}\, \nabla_0 \Sigma\nabla_0 \Sigma^\dagger
  -
  {f_s^2\over4} {\rm Tr}\, \partial_i \Sigma\partial_i \Sigma^\dagger
%  - m_\pi\<\bar\psi\psi\>_{\rm vac} 
  + {m^2 f_s^2\over2}
  {\rm Re }\,{\rm Tr}\, \Sigma \,.
\label{leff:T}
\end{equation}
Due to the lack of Lorentz invariance, $f_t^2$ and $f_s^2$ are
independent parameters.  The covariant derivative $\nabla_0$ is the
same as defined in Eq.\ (\ref{deriv}).  The dispersion relation following
from this Lagrangian has the form (\ref{omega}) where
\begin{equation}
  u = {f_s\over f_t} \, .
  \label{ufsft}
\end{equation}
At zero temperature $f_t=f_s=f_\pi$, and the pion velocity $u$ is equal
to the speed of light.  We now show that, at finite temperature, all
three parameters $f_t$, $f_s$, and $m$ can be determined from equal-time
(or static)
correlation functions.

Repeating the same argument as in Sec.\ \ref{sec:mui5}, we can show
that $f_t$ is related to the axial isospin susceptibility,
\begin{equation}\label{chi_ft}
  \sus=f_t^2 \, ,
\end{equation}
but now $\sus$ is defined as the susceptibility at finite temperature.
Note that, as a susceptibility with respect to a conserved charge,
$\sus$ is free of ultraviolet divergences.

We now need a prescription to compute $f_s$ and $m$ from static
correlation functions.
In order to make the connection, 
we generalize the mass term in Eq.\ (\ref{lquark}),
\begin{equation}
  {\cal L}_{\rm quark} = i\psibar \gamma^\mu D_\mu \psi 
  - (\psibar_L M \psi_R + {\rm H.c.})
  + \muai A^3_0 \,,
\label{lquarkA}
\end{equation}
and regard $M$ as an external {\em field}, $M=M(x)$.  The Lagrangian
(\ref{lquarkA}) possesses the symmetry $\psi_L\to L\psi_L$,
$\psi_R\to R\psi_R$, $M\to LMR^\dagger$, where $L,R\in \SU(2)$.
If $M(x)$ is a slowly varying function of $x$, its effect can be
captured in the effective chiral Lagrangian.  The requirement that the
effective Lagrangian preserves this symmetry fixes the form of its mass
term,
\begin{equation}
  {\cal L}_{\rm eff}= 
  {f_t^2\over4} {\rm Tr}\, \nabla_0 \Sigma\nabla_0 \Sigma^\dagger
  -
  {f_s^2\over4} {\rm Tr}\, \partial_i \Sigma\partial_i \Sigma^\dagger
%  - m_\pi\<\bar\psi\psi\>_{\rm vac} 
%  + {m^2 f^2\over2}
-\frac12\<\psibar\psi\>
  {\rm Re }\,{\rm Tr}\, M^\dagger\Sigma \,.
\label{leff:TA}
\end{equation}
We shall limit ourselves to a particular ansatz of the external field
$M(x)$:
\begin{equation}
  M(x)=m_q\,e^{i\alpha^a(x)\tau^a}\,.
\end{equation}

The second derivative of the partition function with respect to
$\alpha^a$ can be computed in both the microscopic theory
(\ref{lquarkA}) and in the effective theory (\ref{leff:TA}).  In the
microscopic theory, we find
\begin{equation}\label{d2e-micro}
{\delta^2{\cal \ln Z}\over \delta \alpha^a(x)\delta\alpha^b(0)} 
=  m_q^2\<\psibar\psi\>^2\,
\<\varphi^a(x) \varphi^b(0)\>\,,
\end{equation}
where $\varphi^a$ is defined in Eq.\ (\ref{phipsi}).  On the other hand,
from the the effective Lagrangian (\ref{leff}) we find
\begin{equation}\label{d2e-eff}
{\delta^2{\cal \ln Z}\over \delta \alpha^a(x)\delta\alpha^b(0)} 
= m_q^2\<\psibar\psi\>^2\,
\<\phi^a(x) \phi^b(0)\>\,,
\mbox{ where }\quad
\phi^a(x)\equiv {\rm Re }\,{\rm Tr }\, i\tau^a \Sigma(x)/2\,.
%\ T\sum_{q_0=2\pi Tn}\int {d\q\over(2\pi)^3} e^{-iqx}
%  {\delta^{ab}\over f_t^2q_0^2 + f^2(\q^2+m^2)}.
\end{equation}
Comparing Eqs.\ (\ref{d2e-micro}) and (\ref{d2e-eff}), we see that
correlation functions of $\varphi^a(x)$ defined in the microscopic
theory and of $\phi^a(x)$ defined in the effective theory are equal.
This equality should hold for small momenta when the effective theory
is applicable.

The correlation function of $\phi^a(x)$, on the other hand, 
can be calculated by expanding
the effective Lagrangian to second order in $\phi^a$.
%, using $\Sigma=1-i\tau^a\phi^a+
%{\cal O}(\phi^2)$, and inverting the matrix of the quadratic form.
The Matsubara propagator of $\phi^a$ is
\begin{equation}\label{phiphi-time}
  \int\limits_0^{1/T}\!\!d\tau\!\int\!d\x\,
  e^{iq\cdot x}\<\phi^a(x) \phi^b(0)\>=
  {\delta^{ab}\over f_t^2q_0^2 + f_s^2(\q^2+m^2)}\,, \qquad q_0=2\pi Tn\,.
\end{equation}
This propagator, for $q_0=0$ and small $\q$, should be equal to the
propagator of $\varphi^a$.  This establishes Eq.\ (\ref{fm}), with the
identification $f=f_s$.  Together with Eqs (\ref{ufsft}) and (\ref{chi_ft}),
this is our result for the pion dispersion relation.
Furthermore, it is natural to assume that, for small $\q$, the dynamics
of $\varphi^a$ is slow, so at high enough temperature ($T\gg m_\pi$) one
can regard $\varphi^a(\tau,\x)$ as independent of $\tau$.  In this case
the left hand side of Eq.\ (\ref{phiphi-time}) is equal to that of
Eq.\ (\ref{fm-eq}).

It is instructive to write Eq.\ (\ref{fm-eq}) in coordinate space,
\begin{equation}
  \<\varphi^a(0,\x)\varphi^b(0,{\bf 0})\> =
%   {T\over f^2} 
%  \int {d\q\over(2\pi)^3} e^{i\q\cdot\x}
%{\delta^{ab}\over q^2+m^2} 
%  = 
  {T\over4\pi f_s^2} {e^{-m|\x|}\over|\x|}
\quad (|\x|\gg T^{-1}, m_\sigma^{-1})
\,.
  \label{phiphi}
\end{equation}
We see that measuring, in the microscopic theory, the large-distance
equal-time correlation function of the $\varphi^a$ defined in
Eq.\ (\ref{phipsi}) we can extract the screening mass $m$ and the decay
constant $f$.
%the rate of the Yukawa-type exponential fall-off, and the decay constant $f$
%from the magnitude of this correlator.
% It remains to identify $\varphi^a$, defined in Eq.\ (\ref{Sigmaphi}),
% with an operator of QCD.  This operator should transform the same way
% as $\varphi^a$ under chiral rotations.  The obvious candidate the
% operator in Eq.\ (\ref{phipsi}).  Thus, from the equal-time correlator
% of this operator, one can determine $f$ and $m$ according to Eq.\
% (\ref{phiphi}).  
Combined with the determination of the susceptibility
$\sus$ and Eq.\ (\ref{chi_ft}), 
the dispersion relation of soft pions is now completely known.

Below we shall provide a more systematic proof of these relationships
using an operator approach, with crucial inputs from the hydrodynamic
theory.  We shall also demonstrate the validity of the relation
(\ref{u}) in an explicit lowest order perturbative calculation in the
linear sigma model.

\section{Hydrodynamic (operator) approach}
\label{sec:hydro}

From a modern perspective, hydrodynamics is an effective theory
operating at sufficiently large distance and time scales.  (By
``sufficiently large'' normally we mean scales larger than the mean
free path, or the relaxation time.)  As such, it is the most suitable
framework to discuss low-energy degrees of freedom (like pions) at
finite temperatures.  This modern point of view, as opposed to the
view of hydrodynamics as a purely phenomenological description, has
been in existence for a long time \cite{hydro_corr}.  From this
philosophy, it is not surprising that the same hydrodynamic theory
describes systems with very different microscopic dynamics, and that
systems with different symmetries (like normal fluids and superfluids)
correspond to different hydrodynamic theories.  Similarly to the
effective Lagrangian,
hydrodynamic equations can also be viewed as a particular way to satisfy Ward
identities at finite temperatures \cite{Yaffe}.  In our case,
hydrodynamics is nontrivial due to the chiral symmetry breaking
\cite{Sonhydro}.  We will derive the constraints placed by
hydrodynamics on the dynamics in our problem, and show how our results
on the pion dispersion relation follow from there.  Our treatment is
similar, to some degree, to that of Ref.\ \cite{HHPR69}.

\subsection{Basic assumptions of hydrodynamics}

We shall assume that, as one goes sufficiently far into the infrared,
the dynamics of any interacting finite-temperature system can be
described in terms of a finite number of fields, which will be called
 hydrodynamic variables.  To be relevant in the infrared, the
fluctuations of these fields (either thermal fluctuations, or those
due to external sources) should relax arbitrarily slowly.  This
requirement eliminates most of the degrees of freedom, which typically
relax during a relaxation time determined by the microscopic
dynamics.  However, the following fields are obvious hydrodynamic
variables:

\begin{itemize}
\item[(i)] The densities of conserved quantities, including the energy
density $T^{00}$, the momentum density $T^{0i}$, and the densities of
conserved global charges (i.e., the zeroth components of conserved
currents).  These fields cannot relax quickly because of the
conservation laws.  A configuration where charges fluctuate over a
length scale $L$ much larger than the mean free path can relax only by
diffusion, which takes place over a time proportional to $L^2$.  The
relaxation time diverges with the wavelength of the perturbation.
\item[(ii)] The phases of the condensates which break global
symmetries.  At zero temperature the fluctuations of such phases
correspond to Goldstone bosons, whose energy can be arbitrarily small.
At finite temperature below symmetry restoration, one should also expect
the long-wavelength fluctuations of the condensate phases to relax
slowly.
\item[(iii)] Near the critical temperatures of second-order phase
transitions, the order parameters themselves (not just the phases)
should be considered hydrodynamic variables \cite{Hohenberg}.  
For example, it is believed that, in QCD with
two massless flavors, the chiral phase transition is of the second
order, 
where the $\sigma$ meson becomes degenerate with the pions.  Near
$T_c$, hence, $\sigma$ should be included in the hydrodynamic
description.  In contrast to the fields in the categories (i) and
(ii), the rate of relaxation of order parameters is controlled by the
closeness of $T$ to $T_c$, but not by the wavelength of perturbations.%
\footnote{We shall not consider the possibility of Abelian gauge fields
in the Coulomb phase (as opposed to the Higgs phase)
 present in 
magnetohydrodynamics.}
\end{itemize}

If one makes a further  assumption that the fields listed in (i)--(iii)
exhaust all slowly relaxing ones, then the set of hydrodynamic
degrees of freedom is completely fixed once all symmetries of the
theory and the
pattern of symmetry breaking at the given temperature are known.
In this paper we shall limit ourselves to temperatures far away from
any second order phase transition, so order parameters are excluded
from hydrodynamics.  This set of hydrodynamic variables
always contains the energy and momentum densities $T^{00}$ and
$T^{0i}$.  For QCD below the chiral phase transition, one has, 
in addition, the
densities of baryon, isospin, and axial isospin charges, and the
phases of the chiral condensate.

After identifying the hydrodynamic degrees of freedom, one can proceed
in various ways.  One can ask about the the equations of motion that
the hydrodynamic variables obey.  For our case, it is a rather
nontrivial task, because of the multitude of the fields involved.  In Ref.\
\cite{Sonhydro}, the Poisson-bracket technique is used to derive the
dissipationless hydrodynamic equations.  One finds a system of fully
nonlinear coupled differential equations, generalizing the equations
of fluid dynamics and equations of motion of the nonlinear sigma model.
However, the connection of this procedure to the fundamental
(microscopic) field 
theory has not been made, and the physical meaning of several 
temperature-dependent parameters appearing in the final equations 
(called $f_t$, $f_s$, and $f_v$ in Ref.\ \cite{Sonhydro}) is not at all
clear.

Alternatively, one can ask the question: what are the constraints that
hydrodynamics places on the correlation functions of the hydrodynamic
variables?  It is clear from the discussion above that the real-time
correlators of the hydrodynamic variables
[i.e., $\<{\cal O}(t,\x){\cal O}(0,{\bf 0})\>$, 
where ${\cal O}$ is the variable]
are long range, i.e., have
power-law (but not exponential) falloff, at least in the timelike
regime $t\to\infty$, $\x=\mbox{fixed}$.%
\footnote{The hydrodynamic correlators are not required to
have power-law decay in the regime $t=\mbox{fixed}$, $\x\to\infty$.}  
It is less trivial to decide
about the correlators of the operators not belonging to the set of
hydrodynamic variables.  To this end, the following additional 
assumption is made.
\begin{itemize}
\item[(iv)] All local operators can be expressed as local functions of
the hydrodynamic operators and their {\em spatial} derivatives, up to
corrections which have short-ranged correlations that go to zero
exponentially when either temporal or spatial separation goes to infinity 
($t$ or $\x\to\infty$).
%in both timelike and spacelike directions.  
(These short-range parts correspond to the
``noises'' in the hydrodynamic theory.)
\end{itemize}
Physically, these assumptions mean that the dynamics in the infrared
can be described in terms of hydrodynamic variables only,
and is equivalent to the assumption of local thermodynamic equilibrium:
the values of all variables are determined by specifying a few.
The 
assumptions (i)--(iv) form the starting point of our construction of
the hydrodynamic equations.

\subsection{Linearized hydrodynamics for soft pions}

To be less abstract, let us consider QCD in the chiral limit, at a
finite temperature below the chiral phase transitions, and zero chemical
potentials.  The hydrodynamic operators are the energy density
$T^{00}$, the momentum density $T^{0i}$, the baryon density
$\psibar\gamma^0\psi$, the densities of vector and axial isospin
charges,
\begin{equation}
  V_0^a = \psibar\gamma^0{\tau^a\over2}\psi\,, \qquad
  A_0^a = \psibar\gamma^0\gamma^5{\tau^a\over2}\psi \,,
  \label{VA}
\end{equation}
and the pion field, defined as Eq.\ (\ref{phipsi}),
\begin{equation}
  \varphi^a = {i\psibar\gamma^5\tau^a\psi\over\<\psibar\psi\>} \,.
\end{equation}
As we shall see below, to {\em linear} order the dynamics of $A^a_0$ and
$\varphi^a$ is decoupled from other modes.

Let us consider the equation for $A_0^a$.  Infinitesimally close to
the chiral limit, we can derive from the QCD Lagrangian 
the familiar PCAC relation:
\begin{equation}
  \d_\mu A^{a\mu} = m_q\<\psibar\psi\>\varphi^a \,.
  \label{PCAC}
\end{equation}
The left hand side contains, in addition to $A_0^a$, the spatial components of
the axial current $A^{ai}$, which are not hydrodynamic variables.
According to the assumption (iv), we can express $A^{ai}$ as local
functions of the hydrodynamic operators and their spatial derivatives,
plus a short-ranged part.  If we work to leading order in the power of
the fields (which means the linear order), the only ones suitable are
$\varphi^a$ and $A_0^a$ which are parity odd and isovectors.  The
spatial index in $A^{ai}$ forces one to have at least one spatial
derivative.  Thus, to leading orders in fields and derivatives, the
only terms consistent with symmetries are
\begin{equation}
  A^{ai} = - f^2\d_i\varphi^a - D\d_i A_0^a - \xi^{ai} \,,
  \label{Aai}
\end{equation}
where $f^2$ and $D$ are coefficients depending on the temperature, and
$\xi^a_i$ is the short-range (``noise'') part of $A^{ai}$.  (At this
step, we have not yet related $f^2$ to the static correlation functions.
We will do it later on.)
%We will return to the question whether the the expansion (\ref{Aai}) 
%can be trusted later on.  
Equation (\ref{PCAC}) now takes the form
\begin{equation}
  \d_0 A_0^a = f^2\nabla^2\varphi^a + m_q\<\psibar\psi\>\varphi^a
  + D\nabla^2 A_0^a + \d_i\xi^{ai} \, .
  \label{flow1}
\end{equation}
The parameter $D$ can be interpreted as the diffusion coefficient for
$A^a_0$; however, Eq.\ (\ref{flow}) is more complex than a diffusion
equation.  Now let us discuss the equation for $\varphi^a$.  The time
derivative of $\varphi^a$, not being a hydrodynamic operator, can be
expanded as
\begin{equation}
  \d_0\varphi^a = {1\over\chir} A_0^a - \kappa_1'\varphi^a +
       \kappa_2 \nabla^2 \varphi^a + \eta^a \,,
  \label{Josephson1}
\end{equation}
where $\chir$, $\kappa_1'$, and $\kappa_2$ are again coefficients
dependent on temperature, and $\eta$ is a short-range noise.  We have
kept the term with the second derivative of $\varphi^a$ as well as the term
with no derivative.  The reason for doing so is that $\kappa_1'$ is
suppressed by the quark masses.  Indeed, in the chiral limit the state
with $\<\varphi^a\>\neq0$, $\<A_0^a\>=0$ can be another vacuum which
stays unchanged with time.  This is consistent with Eq.\
(\ref{Josephson}) only if $\kappa_1'=0$ in this limit, thus at small
quark masses $\kappa_1'$ is small.  For this reason
we keep the $\nabla^2\varphi^a$ term.

Introducing the parameter $m^2$ defined so that
$f^2m^2=-m_q\<\psibar\psi\>$ (which we still have to relate to static
correlators), and $\kappa_1$ so that $\kappa_1'=\kappa_1m^2$, Eqs.\
(\ref{flow1}) and (\ref{Josephson1}) can be written as
\begin{subequations}\label{flowJos}
\begin{eqnarray}
  \d_0 A_0^a &=& f^2(\nabla^2-m^2)\varphi^a
  + D\nabla^2 A_0^a + \d_i\xi^{ai} \label{flow}\, ,\\
  \d_0\varphi^a &=& {1\over\chir} A_0^a + 
   (\kappa_2\nabla^2-\kappa_1m^2)\varphi^a  + \eta^a \,.
  \label{Josephson}
\end{eqnarray}
\end{subequations}

Equations (\ref{flowJos}) are the linearized
hydrodynamic equations governing the evolution of $A^a_0$ and $\varphi^a$.
The correlation functions of $A^a_0$ and $\varphi^a$
can be found if one knows the correlators of the ``noises'' $\xi^{ai}$
and $\eta^a$.  By construction, these fields have only short-range
correlations, so, if one is interested only in the dynamics at large
distance and time scales, these correlations can be replaced by delta
functions.  Isospin symmetry and rotational invariance require the
correlators to be of the following forms:
\begin{subequations}\label{noises}
\begin{eqnarray}
  \<\xi^{ai}(x) \xi^{bj}(y)\> &=&
     F_\xi \delta^{ab}\delta^{ij}\delta^4(x-y)\,,\\
  \<\eta^a(x)\eta^b(y)\> &=& F_\eta \delta^{ab}\delta^4(x-y)\,,\\
  \<\xi^{ai}(x)\eta^b(y)\> &=& 0 \,.
\end{eqnarray}
\end{subequations}
up to corrections proportional to derivatives of $\delta^4(x-y)$ which
will be neglected since they are of higher order in momentum.
Equations (\ref{flowJos}) and (\ref{noises}) completely determine the
hydrodynamics of soft pions, to the linearized order.

\subsection{Relation to static correlators}

Our next task is to relate, as much as possible, the parameters
appearing in Eqs.\ (\ref{flowJos}) and (\ref{noises}) with the equal-time
correlators of $A^a_0$ and $\varphi^a$.

First, we multiply Eq.\ (\ref{Josephson}) by $A^b$, taken at the
same time moment, and integrate over space.  One finds
\begin{equation}
\begin{split}
  \int\!d\x\, \< \dot\varphi^a(t, \x) A_0^b(t, {\bf 0}) \> 
  =& \int\!d\x\, \left[{1\over\chir}\<A_0^a(t,\x)A_0^b(t,{\bf 0})\>
    -\kappa_1m^2\<\varphi^a(t,\x)A_0^b(t,{\bf 0})\> \right.\\
   & \left. +\<\eta^a(t,\x)A_0^b(t,{\bf 0})\>\right] \,.
\end{split}
  \label{d0phiA}
\end{equation}
In the right hand side, the second term is proportional to $m^2$,
which is small in the chiral limit, and hence can be neglected.  The
last term in the integrand will be shown later to vanish,
\begin{equation}\label{etaA0}
  \<\eta^a(t,\x)A_0^b(t,{\bf 0})\> = 0 \,.
\end{equation}
Equation (\ref{etaA0}) is trivial if
understood as $\<\eta^a(t+\varepsilon,\x)A_0^b(t,{\bf0})\>=0$: 
$A_0^b$ cannot be correlated with the noise $\eta$ in the
future.  What is somewhat less trivial (and will be checked 
{\em a posteriori}) 
is that 
$\<\eta^a(t-\varepsilon,\x)A_0^b(t,{\bf0})\>$ also vanishes: the 
equal-time correlator of $\eta$ and $A_0$ does not depend on how the 
equal-time limit is taken.
Therefore, only the first term survives; and by definition of 
$\sus$ it is equal to
\begin{equation}
  \int\!d\x\, {1\over\chir}\<A_0^a(t,\x)A_0^b(t,{\bf 0})\> 
  = {T\sus\over\chir}\,.
  \label{susoverchi}
\end{equation}

On the other hand, the left hand side of Eq.\ (\ref{d0phiA}) can be
computed explicitly.  To this end, we write
\begin{equation}
  \< \dot\varphi^a(\x)A_0^b({\bf 0})\> = 
  i{\rm Tr}\, \{e^{-\beta H} [H, \varphi^a(\x)]A_0^b({\bf 0})\} \,.
  \label{lhs}
\end{equation}
(we drop the time variable $t$ which is an argument of all operators).
We now make use of the following expansion:
\begin{equation}
  [e^{-\beta H}, \varphi^a] = -\beta e^{-\beta H} [H, \varphi^a]
   - {\beta^2\over2} e^{-\beta H}[H, [H, \varphi^a]] + \cdots\,.
  \label{expansion}
\end{equation}
The expansion parameter here is 
$\beta q_0\equiv q_0/T$ where $q_0$ is the
frequency of variation of $\varphi^a$.  Since we are dealing with the
low-frequency modes in $\varphi^a$, we can ignore all terms beyond the
first in Eq.\ (\ref{expansion}).  Equation (\ref{lhs}) now becomes
\begin{equation}
  \< \dot\varphi^a(\x)A_0^b({\bf 0})\> = 
  -iT\, {\rm Tr}\, \{[e^{-\beta H},\varphi^a(\x)]A_0^b({\bf 0})\}
  = -iT \< [\varphi^a(\x), A_0^b({\bf 0})]\> = T\delta^{ab}\delta^3(\x) \,.
\end{equation}
In the last transformation, we make use of the commutation relation
\begin{equation}
  [\varphi^a(\x), A_0^b({\bf 0})] = i\delta^{ab}\delta^3(\x) 
  {\psibar\psi\over\<\psibar\psi\>} \,.
\end{equation}
%(NOTE: CORRECT ONLY WHEN $N_f=2$).
Comparing to Eq.\ (\ref{susoverchi}), we find
\begin{equation}
  \chir = \sus\, .
\end{equation}
Thus, we show that $\chir$ is the axial isospin susceptibility of
the system, which is a static quantity.  The proof we have just
presented is similar to that of the equipartition theorem in
statistical mechanics.

Analogously, we can show that
\begin{equation}
  \int\!d\x\, e^{-i\q\cdot\x}
  \<\dot A_0^a(\x) \varphi^b({\bf 0})\> = -T\delta^{ab} \,.
  \label{A0phicomm}
\end{equation}
On the other hand, from Eq.\ (\ref{flow})
\begin{equation}
\begin{split}
  \int\!d\x\, e^{-i\q\cdot\x} \<\dot A_0^a(\x) \varphi^b({\bf 0})\> =& 
  -f^2 (\q^2 + m^2) 
  \int\!d\x\, e^{-i\q\cdot\x} \<\varphi^a(\x)\varphi^b({\bf 0})\> \\
  & - D\q^2 \int\!d\x\, e^{-i\q\cdot\x} \<A_0^a(\x)\varphi^b({\bf 0})\> \,,
\end{split}
  \label{A0phiflow}
\end{equation}
where we dropped the term proportional to
\begin{equation}\label{xiphi}
  \<\d_i\xi_i^a(t,\x)\varphi^b(t,{\bf 0})\> 
\end{equation}
as it will be shown to vanish
in the same manner as $\<\eta^a A_0^b\>$ in Eq.\ (\ref{d0phiA}).
Moreover, $Dq^2\<A_0^a\varphi^b\>$ can be neglected compared to
$\<\dot A_0^a\varphi^b\>\sim q_0\<A_0^a\varphi^b\>$, since we expect
the pion to have a linear dispersion relation and, for small enough
momenta, $q_0\gg Dq^2$.  Equating Eqs.\ (\ref{A0phicomm}) and
(\ref{A0phiflow}), we find that equal-time correlators of the pion
field must have the form of a Yukawa potential,
\begin{equation}
  \int\!d\x\,e^{-i\q\cdot\x} \<\varphi^a(\x) \varphi^b({\bf 0})\> = 
  {T\over f^2} {\delta^{ab}\over \q^2+m^2} \,.
  \label{phiphiYukawa}
\end{equation}
Thus, the parameters $f^2$ and $m^2$ appearing in Eq.\ (\ref{flow}) are
the same ones defined in Eq.\ (\ref{fm-eq}) via the equal-time correlator
of $\varphi^a$.

The parameters $D$, $\kappa_1$, $\kappa_2$, $F_\xi$, and $F_\eta$
cannot be expressed individually in terms of the equal-time
correlators, but some relations between them will be derived below.

\subsection{Hydrodynamic correlation functions}

From Eqs.\ (\ref{flowJos}) and (\ref{noises}) one can easily compute the
real-time correlators of $A^a_0$ and $\varphi^a$:
\begin{subequations}\label{hydrocorr}
\begin{eqnarray}
  \int\!d^4x\, e^{iq\cdot x}\, \<A^a_0(x) A^b_0(0)\> &=& \delta^{ab}
  {q_0^2\q^2 F_\xi + \sus^2 \omega_\q^4 F_\eta \over
  [(q_0-\omega_\q)^2+{1\over4}\Gamma_\q^2]
  [(q_0+\omega_\q)^2+{1\over4}\Gamma_\q^2]}\,,
  \label{hydrocorrA}\\
  \int\!d^4x\, e^{iq\cdot x}\, \<\varphi^a(x) \varphi^b(0)\> &=& 
  \delta^{ab} 
  {\sus^{-2}\q^2 F_\xi + q_0^2 F_\eta \over
  [(q_0-\omega_\q)^2+{1\over4}\Gamma_\q^2]
  [(q_0+\omega_\q)^2+{1\over4}\Gamma_\q^2]}
  \, ,
  \label{hydrocorrphi}
\end{eqnarray}
\end{subequations}
where
\begin{subequations}\label{disp}
\begin{eqnarray}
  \omega_\q^2 &=& {f^2\over\sus}(\q^2+m^2) \label{disp-re}\,,\\
  \Gamma_\q &=& \kappa_1 m^2 + (D+\kappa_2)\q^2\,. \label{disp-im}
\end{eqnarray}
\end{subequations}
The correlation functions (\ref{hydrocorr})
peak around $q_0\approx\pm\omega_\q$, and the width of the peaks
$\Gamma_\q$ is much smaller than $\omega_\q$ at small $\q$.  The
correlators have poles corresponding to the pion collective
excitations with the dispersion relation
$q_0=\omega_\q-{i\over2}\Gamma_\q$.  We note, moreover, that all
parameters entering the real part of the dispersion relation
(\ref{disp-re}) can be determined from the static correlation
functions.

It is now possible to explicitly check that the noise-field
correlators appearing in Eqs.\ (\ref{etaA0}) and (\ref{xiphi}) vanish.
Indeed, by using Eqs.\ (\ref{flowJos}) and (\ref{noises}) we find
\begin{subequations}\label{noisefield}
\begin{eqnarray}
  \<\eta(t,\x)A_0(t',{\bf 0})\> &=& \delta^{ab}\!\int\!{d^4q\over(2\pi)^4}\,
  e^{-iq_0(t-t')+i\q\cdot\x}\, {\sus\omega_\q^2F_\eta\over
  q_0^2-\omega_\q^2+iq_0\Gamma_\q} \,,\\
  \<\d_i\xi_i(t,\x)\varphi(t',{\bf 0})\> &=& 
  -\delta^{ab}\!\int\!{d^4q\over(2\pi)^4}\,
  e^{-iq_0(t-t')+i\q\cdot\x}\, {\sus^{-1}\q^2F_\xi\over
  q_0^2-\omega_\q^2+iq_0\Gamma_\q} \,.
\end{eqnarray}
\end{subequations}
The integrands in Eqs.\ (\ref{noisefield}) have two poles, both
located in the upper half plane: $q_0=\pm\omega_\q+{i\over2}\Gamma_q$.
When $t>t'$, when taking integrals over $q_0$, one can close the
contour in the lower half plane, so the integrals are obviously zero,
which corresponds to our previous remark that the fields cannot
correlate with the noises in the future.  If one takes $t\to t'$ from
below, the integrals are also zero since they are equal to the sums of
the residues of the integrands (which are zero because these functions
behave as $q_0^{-2}$ at large $q_0$.)  Therefore, we have checked
that the equal-time noise-field correlators (\ref{etaA0}) and
(\ref{xiphi}) indeed vanish.

One can find the relations between the amplitude of the noise
correlators, $F_\xi$ and $F_\eta$, and the parameters characterizing
the damping in Eq.\ (\ref{disp-im}).  One integrates Eqs.\
(\ref{hydrocorr}) over $q_0$ and obtains the
equal-time correlation functions
\begin{subequations}\label{Aphicorr}
\begin{eqnarray}
  \int\!d\x\, e^{-i\q\cdot\x} \< A^a_0(t,\x) A^b_0(t,{\bf 0}) \> &=&
  \delta^{ab} {\q^2F_\xi+\sus^2\omega_\q^2 F_\eta\over2\Gamma_\q}\,,\\
  \int\!d\x\, e^{-i\q\cdot\x} 
    \< \varphi^a(t,\x) \varphi^b(t,{\bf 0}) \> &=&
  \delta^{ab} {\q^2F_\xi+\sus^2\omega_\q^2F_\eta
    \over 2\sus^2 \omega_\q^2 \Gamma_\q}\,.
\end{eqnarray}
\end{subequations}
Comparing these correlators with Eqs.\ (\ref{chi-eq}) and
(\ref{fm-eq}), taking into account Eqs.\ (\ref{disp}), one finds
\begin{subequations}\label{F}
\begin{eqnarray}
  F_\xi &=& 2T\sus(D+\kappa_2-\kappa_1) \,,\\
  F_\eta &=& {2T\kappa_1\over f^2}  \,. 
\end{eqnarray}
\end{subequations}
Our results for the hydrodynamic correlations functions are given by
Eqs.\ (\ref{hydrocorr}) and (\ref{F}).
%\begin{eqnarray}
%  \int\!d^4x\, e^{iq\cdot x}\, \<A^a_0(x) A^b_0(0)\> &=& 
%  {T\delta^{ab}\sus\Gamma_\q\delta^{ab}\over2} \left[
%  {1\over (q_0{-}\omega_\q)^2 {+} {1\over4}\Gamma_\q^2} +
%  {1\over (q_0{+}\omega_\q)^2 {+} {1\over4}\Gamma_\q^2} 
%  \right] , \nonumber\\
%  \int\!d^4x\, e^{iq\cdot x}\, \<\varphi^a_0(x) \varphi^b_0(0)\> &=& 
%  {T\Gamma_\q\delta^{ab} \over 2\sus\omega_\q^2}
%  \left[
%  {1\over (q_0{-}\omega_\q)^2 {+} {1\over4}\Gamma_\q^2} +
%  {1\over (q_0{+}\omega_\q)^2 {+} {1\over4}\Gamma_\q^2} 
%  \right] .
%  \label{hydrocorr}
%\end{eqnarray}
For completeness, we give here also the result for the cross
correlator of $\varphi^a$ and $A_0^a$:
\begin{equation}
  \int\!d^4x\, e^{iq\cdot x}\,
  \<\varphi^a(x)A^b_0(0)\> = {2iTq_0\Gamma_\q\over
  [(q_0-\omega_\q)^2+{1\over4}\Gamma_\q^2]
  [(q_0+\omega_\q)^2+{1\over4}\Gamma_\q^2]} \, .
\end{equation}

It is instructive to compare the correlator of the axial isospin
charge density $A^a_0$ [Eq.\ (\ref{hydrocorrA})] with that of the
vector isospin charge density $V^a_0$.  The dynamics of the latter is
completely diffusive and is given by the equations
\begin{subequations}
\begin{eqnarray}
  \d_0 V^a_0 - \Di\nabla^2 V^a_0 &=& \d_i \zeta^{ai} \, ,\\
  \< \zeta^{ai}(x) \zeta^{bj}(0) \> &=& 
  2T\Di\susi\delta^{ab}\delta^{ij}\delta^4(x)\,,
\end{eqnarray}
\end{subequations}
where $\Di$ is the diffusion constant for the isospin charge, and
$\susi$ is the (vector) isospin susceptibility.  The correlator of
$V^a_0$ is
\begin{equation}
  \int\!d^4x\, e^{iq\cdot x} \< V^a_0(x) V^b_0(0) \> =
  {2T\Di\susi \q^2\over q_0^2 + \Di^2\q^4}\, .
  \label{hydrocorrV}
\end{equation}
The pole of the correlator is located at purely imaginary frequencies
$q_0=\pm i\Di\q^2$, as it should be for a purely diffusive mode.

Using Eqs.\ (\ref{hydrocorrV}) and (\ref{Aphicorr})
we can understand the limitations of the Lagrangian approach
and the role of the dissipative processes.
For example, one could try to apply the method
of Sec.~\ref{sec:lagr} to determine the vector
isospin susceptibility $\susi$, introducing an isospin chemical potential
$\mu_I$. At zero temperature 
such a method was used in Ref.\ \cite{mui}. 
%The result is $\susi=0$. 
As expected, the isospin susceptibility $\susi$
vanishes at $T=0$ since it takes
finite energy to excite isospin degrees of freedom (pions)
and change the isospin density.
However, at {\em nonzero} temperature, 
the naive application of the effective
Lagrangian method would predict that $\susi=0$ also.%
\footnote{We thank T. Sch\"afer for pointing this out to us.}
On the other hand, we should expect $\susi\neq 0$ at finite temperature,
even at very small $T$, 
because of the presence of isospin-carrying pions in the thermal medium.
Looking at the correlator (\ref{hydrocorrV}), one sees that the equal-time 
correlator defining  $\susi$,
being the integral of Eq.\ (\ref{hydrocorrV}) over $q_0$ in the limit 
$\q\to0$, 
receives the main contribution from very slow (diffusive) 
modes: $q_0\sim D_I \q^2$. However, the effective Lagrangian (\ref{leff:T})
only describes faster (propagating) modes of the $\varphi$ field
with $q_0=\omega_\q \gg D_I\q^2$. The slow diffusive modes which
contribute to
$\susi$ are not present in the Lagrangian (\ref{leff:T}).
For the axial isospin susceptibility $\sus$, the situation is
different, for the integral of Eq.\ (\ref{hydrocorrV}) 
over frequencies in the limit $\q\to 0$
is concentrated entirely near values $q_0=\pm\omega_\q$. Thus, unlike
$\susi$, the value
of $\sus$ at finite $T$ 
can be found correctly using the effective Lagrangian approach.

\subsection{\label{sec:hydro-comp}Comparison to previous results}

As a by-product of our analysis, we obtain a generalization of the
Gell-Mann--Oakes--Renner relation to finite temperature:
\begin{equation}
  f^2 m^2 = -m_q \< \psibar\psi \> \,.
  \label{GORscr}
\end{equation}
This equation has the same form as at zero temperature.  At finite
temperature, $m$ should be understood as the pion screening mass, and
the exact meaning of the temperature-dependent ``pion decay constant''
$f^2$ is given by Eq.\ (\ref{phiphiYukawa}).  One can also write the
GOR relation in an alternative form,
\begin{equation}
  \sus m_p^2 = -m_q \< \psibar\psi \> \,,
  \label{GORpole}
\end{equation}
where $m_p=um$ is the ``pole mass'' of pions.

That the velocity of pions at finite temperature is different from the
velocity of light has been seen in the second order of chiral
perturbation theory (i.e., in order $T^4/f_\pi^4$) \cite{Pisarski,Toublan}.
The authors of Ref.\ \cite{Pisarski} also introduced two pion decay
constants $f^t$ and $f^s$, which correspond to our $\sqrt{\sus}$ and $f$,
and checked the validity of the GOR relation (\ref{GORpole}).
However, these constants were defined only in chiral perturbation
theory, and only at small $T$; there has not been any attempt to give
a precise definition of the constants $f^t$ and $f^s$ at temperatures
comparable to $T_c$.  We have, in contrast, given a precise meaning
to the constants $\sus$ and $f$ in terms of equal-time correlation
functions that can be measured on the lattice.  The GOR relation now
contains only well-defined quantities.  Although Eq.\ (\ref{GORpole})
cannot be checked on the lattice, since it contains the pion {\em pole}
mass, the version (\ref{GORscr}) can be verified numerically since all
quantities entering it are statically measurable.

We also note that Eqs.\ (\ref{flowJos}), without
the noise and dissipation terms, can be obtained by linearizing the
hydrodynamic equations obtained in Ref.\ \cite{Sonhydro} by the
Poisson-bracket technique.  One can then identify the parameters
$f_t$ and $f_s$ in Ref.\ \cite{Sonhydro} as $f_t^2=\sus$, $f_s^2=f^2$
(the parameter $f_v^2$ of Ref.\ \cite{Sonhydro}) is equal to the
vector isospin susceptibility).  We did not try to reproduce the full
nonlinear hydrodynamic equation of Ref.\ \cite{Sonhydro} in the
present approach.

\section{An explicit example: a linear sigma model}
\label{sec:scalar}

It is instructive to explicitly verify our relation (\ref{u})
between the
velocity of the Goldstone bosons, the axial isospin susceptibility,
and the temperature-dependent decay constant in a model where
weak-coupling calculations are possible.  The simplest theory with
spontaneous breaking of a continuous symmetry is the linear sigma
model.  
%We shall first discuss the linear sigma model without fermion
%where we show explicitly that Eq.\ (\ref{u}) is valid to leading
%order of perturbation theory.  When fermions are added into the theory,
%it is no longer sufficient to limit oneself to a finite number of 
%Feynman diagrams, so the computation is much more complicated.  We do
%not present here the complete resummation of Feynman graphs but
%only indicate the physical arguments showing that one should recover
%the formula (\ref{u}) when such a resummation is performed.
We shall show that Eq.\ (\ref{u}) indeed holds in this model to leading
order of perturbation theory

It is important to emphasize that the sigma model considered in this
section is only meant to serve as an example of a theory where we can
explicitly check our results by perturbative calculations.  These
models do not describe QCD at $T$ of order $T_c$ --- a theory
with no apparent small parameter.  We claim, however, that the
connection between the dispersion relation of the pions and static
correlation functions is model independent, and can be derived from
only a few general assumptions stated in Sec.\ \ref{sec:hydro}.

%\subsection{Linear sigma model without fermions}

We start with the following Lagrangian
\begin{equation}
  {\cal L} =  {1\over2}\d^\mu\phi^a\d_\mu\phi^a +
      {\mu^2\over2} \phi^a\phi^a 
      -{\lambda\over4} (\phi^a\phi^a)^2 \,,
  \label{L}
\end{equation}
where $a=1,2,\ldots,N$.  We assume $\lambda\ll1$, so perturbation
theory can be applied.  When $\mu^2>0$, the $O(N)$ symmetry is
spontaneously broken at zero temperature.  We choose the vacuum to
align in the $\phi^N$ direction.  At zero temperature, and at tree
level, the vacuum expectation value of $\phi^N$ is
\begin{equation}
  \<\phi^N\> \equiv v_0 = {\mu^2\over\lambda} \,.
\end{equation}
At finite temperature the expectation value of $\phi^N$ is different
from $v_0$.  Denoting it as $v$ and replacing $\phi^N=v+\sigma$ in the
Lagrangian (\ref{L}), we obtain
\begin{equation}\label{Lexpand}
\begin{split}
  {\cal L} &= {1\over2}(\d_\mu\sigma)^2 + {1\over2}(\d_\mu\bfpi)^2
      - (\lambda v^2 {-} \mu^2) v\sigma
      - {1\over2}(3\lambda v^2{-}\mu^2)\sigma^2 
      - {1\over2}(\lambda v^2 {-} \mu^2)\bfpi^2 \\
   & \quad-\lambda v\sigma^3 -\lambda v\sigma\bfpi^2
      - {\lambda\over4}\sigma^4 - {\lambda\over4}(\bfpi^2)^2
      - {\lambda\over2}\sigma^2\bfpi^2 \,,
\end{split}
\end{equation}
where $\bfpi=(\phi^1,\phi^2,\ldots,\phi^{N-1})$.  From Eq.\
(\ref{Lexpand}) the Feynman rules can be easily written down.

The value of $v$ is determined by the condition that $\<\sigma\>=0$.
For computation, we will use the Matsubara (Euclidean) formalism, in
which this condition reads, to one-loop order,
\begin{equation}
  (\lambda v^2 -\mu^2)v + \lambda vT\sum_{p_0}\int\!{d\p\over(2\pi)^3}\,
  \biggl( {3\over P^2+m_\sigma^2} 
  + {N-1\over P^2}\biggr) = 0 \,,
  \label{tadpole}
\end{equation}
In this Section, $P^2=p_0^2+p^2$, $p=|\p|$.  The sum-integral in
Eq.\ (\ref{tadpole}) is ultraviolet divergent.  This
temperature-independent divergence can be
absorbed into the redefinition of $\mu^2$.  The temperature dependence
of $v$ comes from the thermal part of the Lagrangian.  We shall be
interested in temperatures of order $v_0$, so $T\gg m_\sigma$.  In
this case,
\begin{equation}
  \lambda v^2  - \mu^2 + {N+2\over12}\lambda T^2 = 0 \,,
\end{equation}
or
\begin{equation}
  v^2(T) = v_0^2 \biggl(1-{T^2\over T_c^2}\biggr)\, \quad\mbox{ with } \quad
  T_c^2 = {12\over N+2} {\mu^2 \over \lambda} \,.
  \label{vT}
\end{equation}
Equation (\ref{vT}) is valid everywhere except for a narrow Ginzburg region
near $T_c$.

The masses of $\sigma$ and $\pi$ are computed, e.g., in Ref.\
\cite{Kapusta}.  At the one-loop level $m_\sigma$ receives
contributions from four diagrams
$$
  \def\epsfsize #1#2{0.7#1}
  \epsfbox{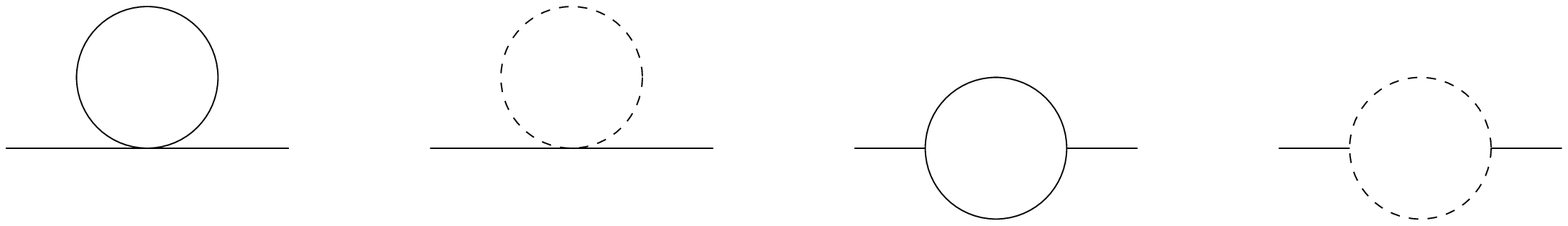}
$$
(where $\sigma$ propagators are denoted by solid lines, and $\pi$
propagators are drawn as dashed lines).  The last two bubble diagrams
are negligible in the finite-temperature regime we consider, so
\begin{equation}
  m_\sigma^2 = 3\lambda v^2 - \mu^2 + 
  \lambda vT\sum_{p_0}\int\!{d\p\over(2\pi)^3}\,
  \biggl( {3\over P^2+m_\sigma^2} + {N-1\over P^2}\biggr) 
  = 2\lambda v^2 \,,
\end{equation}
and thus decreases with temperature.  The Goldstone boson receives
corrections to its mass from three one-loop graphs,
$$
  \def\epsfsize #1#2{0.7#1}
  \epsfbox{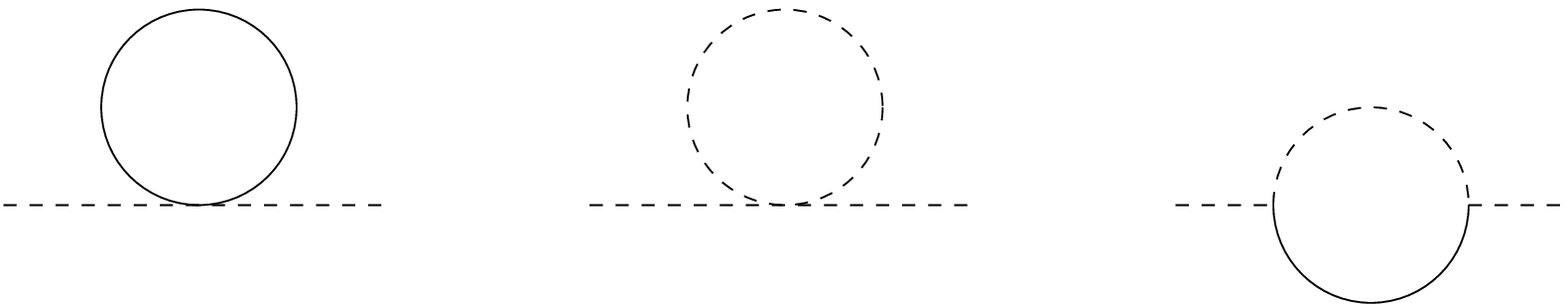}
$$
and remains massless since
\begin{equation}
\begin{split}
  m_\pi^2 &= \lambda v^2\mu^2 - \mu^2 + \lambda v T\sum_{p_0}\int\!
  {d\p\over(2\pi)^3}\, \biggl( {1\over P^2+m_\sigma^2} +
  {N-1 \over P^2}\biggr)\\
   & \quad - 4\lambda^2 v^2 T\sum_{p_0} \int\!{d\p\over(2\pi)^3}\,
  {1\over P^2(P^2+m_\sigma^2)} = 0 \,.
\end{split}
\end{equation}

To find the dispersion relation of the pion near $q=0$, one has to
expand the pion self-energy $\Sigma(q)$ in powers of $q_0$ and $\q$.
The only one-loop diagram that depends on $q$ is the bubble diagram,
which can be evaluated either by using the Schwinger-Keldysh real-time
formalism, or by doing calculations in the Matsubara formalism and then
performing an analytic continuation from imaginary to real $q_0$.  By
either method one finds
\begin{eqnarray}
  \Sigma(q) &=& -4\lambda^2 v^2 \int\!{d\p\over(2\pi)^3
  2\omega_{\p+\q}2\Omega_\p}\,  \biggl\{
  [f(\omega_{\p+\q}){+}f(\Omega_\p){+}1]
  \biggl({1\over\Omega_\p{+}\omega_{\p+\q}{-}q_0} 
  + {1\over\Omega_\p{+}\omega_{\p+\q}{+}q_0}\biggr) \nonumber\\
  & & + [f(\omega_{\p+\q}){-}f(\Omega_\p)]\biggl(
  {1 \over \Omega_\p {-} \omega_{\p+\q} {-} q_0} +
  {1 \over \Omega_\p {-} \omega_{\p+\q} {+} q_0}\biggr)\biggr\} \,,
  \label{Sigma}
\end{eqnarray}
where $\omega_\p$ and $\Omega_\p$ are the energies of the $\pi$ and
$\sigma$ particles with momentum $\p$: $\omega_\p=p$,
$\Omega_\p=(p^2+m_\sigma^2)^{1/2}$; and $f(\omega)$ is the
Bose-Einstein distribution function
$f(\omega)=(e^{\beta\omega}-1)^{-1}$.

We need to compute the coefficients of $q_0^2$ and $q^2$ in the
expansion of $\Sigma(Q)$.  To compute the coefficient of $q^2$, one
can set $q_0=0$ and Eq.\ (\ref{Sigma}) becomes
\begin{equation}
  \Sigma(0, \q) = 4\lambda^2 v^2 \int\!{d\p\over(2\pi)^3}\,
  {1\over\Omega_\p^2-\omega_{\p+\q}^2} \Biggl[
  {f(\Omega_\p)+{1\over2}\over\Omega_\p} -
  {f(\omega_{\p+\q})+{1\over2}\over\omega_{\p+\q}}\Biggr] \,.
  \label{Sigma0q}
\end{equation}
The right hand side of Eq.\ (\ref{Sigma0q}) can be expanded in powers
of $q$.  The constant is compensated by the other diagrams.  The $q^2$
term gives rise to an integral which is dominated in the infrared
region $p\sim m_\sigma$, and is of order
\begin{equation}
  \Sigma(0, \q) \sim {\lambda^2 v^2 T\over m_\sigma^3} \q^2 =
  O(\lambda^{1/2})\,\q^2 + \cdots \,.
\end{equation}
In the last equation we assumed $T\sim v_0$, and
$m_\sigma^2\sim\lambda T$, which is valid when $T$ is not very close
to $T_c$.

Now let us put $\q=0$ and expand in $q_0$.  Since the constant
term is canceled out by other diagrams, one has to look only at terms
of order $q_0^2$.  One finds
\begin{equation}
  \Sigma(q_0, {\bf 0}) = -8\lambda^2 v^2q_0^2\int\!{d\p\over(2\pi)^3
  2\omega_\p2\Omega_\p}\,\Biggl[
  {f(\omega_\p)+f(\Omega_\p)+1\over (\Omega_\p+\omega_\p)^3} +
  {f(\omega_\p)-f(\Omega_\p)\over(\Omega_\p-\omega_\p)^3} \Biggr] \,.
\end{equation}
The first term in the square brackets gives rise to an integral that
is dominated by the infrared, i.e., by $p\sim m_\sigma$, and so is
completely analogous to the coefficient of $p^2$ above.  The
coefficient of $q_0^2$ coming from the first term is hence of order
$O(\lambda^{1/2})$.  In contrast, the integral of the second term in
the square brackets is dominated by $p\sim T$.  For such $p$ one can
write, approximately,
\begin{equation}
  \Omega_\p - \omega_\p = {m_\sigma^2\over2p} \, ,\qquad
  f(\omega_\p) - f(\Omega_\p) = 
    -{m_\sigma^2\over2p} {\d n(p)\over \d p} \,,
\end{equation}
which gives
\begin{equation}
  \Sigma(q_0, {\bf 0}) = {8\lambda^2 v^2\over m_\sigma^4} q_0^2
  \int\!{d\p\over(2\pi)^3}\, {\d f(p)\over\d p} =
  - {4\lambda^2 v^2\over3 m_\sigma^4} T^2 q_0^2 \,.
\end{equation}
Substituting $m_\sigma^2=2\lambda v^2$, we finally find
\begin{equation}
  \Sigma(q_0, \q) = -{T^2\over3v^2}q_0^2 \,.
\end{equation}
Since $T$ and $v$ are both of order $v_0$, the coefficient in front of
$q_0^2$ is of order ${\cal O}(\lambda^0)$, 
in contrast to that of $\q^2$.  The velocity
of pions in our model is
\begin{equation}
  u^2 = \biggl(1+{T^2\over3v^2}\biggr)^{-1}, \qquad T\gg m_\sigma\,.
  \label{u2ON}
\end{equation}
It is instructive to compare this result to the one obtained in the
framework of the chiral perturbation theory: $u=1-{\cal
O}(T^4/f_\pi^4)$
\cite{Pisarski,Toublan}.
One should bear in mind that the regime in which the chiral
perturbation theory result applies corresponds to $T\ll m_\sigma$.
In our linear sigma model we consider a different regime:
$T\sim v \gg m_\sigma\sim \sqrt\lambda v$. In the chiral perturbation
theory $T/f_\pi$ serves as an expansion parameter. In our weak-coupling
calculation the expansion is in $\lambda$ while the $T$ dependence is
included to all orders in $T/v$.

Now we need to check that this coincides with $f^2/\sus$.  Recall that
our definition of $f^2$ is as follows: if we define
$\varphi^a=\pi^a/v$, then
\begin{equation}
  \int\!d\x\, e^{-i\q\cdot \x} \<\varphi^a(t,\x)\varphi^b(t,{\bf 0})\>
  = {T\over f^2}{\delta^{ab}\over \q^2} \,.
\end{equation}
The left hand side is proportional to the $\pi$ Matsubara propagator,
summed over $q_0$,
\begin{equation}
  {T\over v^2}\sum_{q_0}{1\over q_0^2+\q^2} \,.
\end{equation}
When $|\q|\ll T$ the dominant term in the sum is the one with $q_0=0$,
therefore,
\begin{equation}
  f^2=v^2 \,.
  \label{f2ON}
\end{equation}
Taking into account one-loop graphs, as we have seen, will change
$f^2$ only by an amount of order $O(\lambda^{1/2})$.

Now to compute the susceptibility $\sus$ we need to turn on a chemical
potential coupled to a broken charge.  There are $N-1$ broken
generators.  Let us consider the one that transforms $\sigma$ and
$\pi_1$ into each other.  The change of the Lagrangian when the
corresponding chemical potential is turned on is
\begin{equation}
  \delta{\cal L} = \mu(\pi_1\d_0\sigma-\sigma\d_0\pi_1) +
  {\mu^2\over2}[(v+\sigma)^2+\pi_1^2] \,.
\end{equation}

The susceptibility can be computed in Matsubara formalism.  There are
two contributions: one from the $\mu^2$ term in $\delta{\cal L}$, the
other from the bubble graph:
\begin{equation}\label{chiON}
\begin{split}
  \sus &= v^2 + 2T\sum_{p_0}\int\!{d\p\over(2\pi)^3}\,
  {1 \over p_0^2+p^2}
  - 4T\sum_{p_0}\int\!{d\p\over(2\pi)^3}\,{p_0^2\over(p_0^2+p^2)^2}\\
%\end{equation}
%Using the formulas
%\begin{eqnarray}
%  T\sum_{p_0} {1\over p_0^2+p^2} &=& {1\over2p}\coth{p\over 2T}\\
%  T\sum_{p_0} {1\over(p_0^2+p^2)^2} &=& 
%    {1\over4p^3}\coth{p\over2T}+ {1\over8p^2T}\sinh^{-2}{p\over2T}
%\end{eqnarray}
%we find
%\begin{equation}
%  \sus 
  &= v^2 + {1\over2T}\int\!{d\p\over(2\pi)^3}\, \sinh^{-2}
  {p\over2T} = v^2 + {T^2\over3} \,.
\end{split}
\end{equation}
The interpretation of this formula is rather direct: $v^2$ is the
contribution of the condensate, and $T^2/3$ is the contribution from
the free gas of $\sigma$ and $\pi_1$.  The square of the velocity
of pions, Eq.\ (\ref{u2ON}), is equal to the ratio of $f^2$ in Eq.\
(\ref{f2ON}) and $\sus$ in Eq.\ (\ref{chiON}), which is what we need
to verify.  We did not, however, attempted to turn on an explicit
symmetry breaking and verify, e.g., the GOR relation at finite
temperature.  Such a calculation should be straightforward.

\section{Conclusion}
\label{sec:conclusions}

Our goal has been to demonstrate that the dispersion relation of pions
can be expressed in terms of quantities obtainable from equal-time or static
correlation functions.  The precise relation is given in Sec.\
\ref{sec:summary}.  Our result enables one to find the real part of
the pion dispersion relation on the lattice.  However, it does not
enable one to compute the imaginary part, which characterizes the
damping of pion modes.

Nowhere in our treatment did we assume any condition on the
temperature (except $T<T_c$), 
as long as we stay infinitesimally close to the chiral
limit.   Thus, if $m_q$ is infinitesimally small, our result
applies to all temperatures smaller than the temperature of the chiral
phase transition $T_c$.  However, at any fixed (small) value of $m_q$,
our results do not apply at 
%very low temperatures, as well as for
temperatures too close to critical,
%The lower limit of $T$ comes from
%the requirement that, for example, in Eq.\ (\ref{phiphi}) only zero
%frequency contributes.  This implies $2\pi T\ge m_\pi$.  
%At
%temperatures very close to $T_c$, 
where the pion screening mass becomes
of the same order as the screening mass of the sigma meson.  
The width of this region near $T_c$ shrinks to zero as a power of
$m_q$ (more precisely, as $m_q^{1/\beta\delta}$).
Our
treatment must fail there because, as explained in Sec.\ \ref{sec:hydro}, the
sigma boson also needs to be included into the hydrodynamic theory.
However, the scaling of different quantities in this temperature region
can be determined using scaling and universality arguments, 
as discussed in Ref. \cite{critical_pions}. In the Appendix of the
present paper we derive some additional interesting 
scaling properties omitted in Ref. \cite{critical_pions}.

Finally, assuming $m_q$ to be very small, as $T\to T_c$, the chiral condensate
$\<\psibar\psi\>\to0$.  As shown in Ref.\ \cite{critical_pions}, this
implies that also $f\to0$ (although with a slightly different critical
exponent).  On the other hand, the axial isospin susceptibility $\sus$
becomes degenerate with the (vector) isospin susceptibility at $T_c$,
where both remains finite.  One concludes from Eq.\ (\ref{u}) that the
pion velocity $u$ tends to zero as one approaches the critical
temperature.  In fact, it can be shown that $u$ approaches zero faster
than the divergence of the screening mass $m$, so the pion pole mass
$m_p=um$ goes to zero as $T\to T_c$ \cite{critical_pions}.

\begin{acknowledgments}
 We thank Daniel Boyanovsky, Sangyong Jeon, Guy Moore,
Thomas Sch\"afer, Edward Shuryak and Larry Yaffe for stimulating discussions.
The authors thank RIKEN, Brookhaven National Laboratory, and the U.S.\
Department of Energy (DE-AC02-98CH10886) for providing the facilities
essential for the completion of this work.  
D.T.S. is supported, in part, by DOE         
Grant No.\ DOE-ER-41132 and by the Alfred P.~Sloan Foundation.  
M.A.S. is supported by a DOE OJI grant.
\end{acknowledgments}

\appendix

%\section{Nonlinear terms in equations of motions}

\section{Dynamical scaling at the chiral transition}
\label{sec:z}

In this appendix we provide a simple derivation of the dynamical
critical exponent $z$ characterizing critical slowing down
near the QCD chiral phase transition. It is based on a scaling argument
similar to the one used in Ref.\ \cite{Hohenberg} to find
$z$ in antiferromagnets. 

For simplicity, here we consider the chiral limit $m_q=0$.
Scaling and universality arguments presented in Ref.\ \cite{critical_pions}
predict that the velocity $u$ vanishes as $T$ approaches $T_c$
from below. The quantity $u^2$ scales as $u^2\sim t^{(d-2)\nu}$, where
$t\equiv(T_c-T)/T_c$. Since the inverse correlation length of the
order parameter $\psibar\psi$, i.e., the static
screening mass of the sigma particle $m_\sigma$,
scales as $m_\sigma\sim t^\nu$,
we conclude that for $t\ll1$
\begin{equation}\label{umsigma}
u^2\sim m_\sigma^{d-2}\,.
\end{equation}

The fact that $u\to0$ at $T_c$ means that the dispersion relation
ceases to be linear. Moreover, the effect of damping also
becomes important; in other words, we expect the real and imaginary parts
of $\omega$ to become comparable. In such a situation on cannot
refer to $\omega$ as a quasiparticle energy. Rather, it is a 
characteristic frequency, or inverse relaxation time, of a mode of
a given wave number $\p$. Scaling hypothesis dictates that the
relation between $\p$ and $\omega$ should be homogeneous, i.e.,
$\omega\sim |\p|^z$. The dynamical scaling exponent $z$ is,
in a generic case, new exponent independent of the static exponents,
e.g., $\nu$ and $\eta$. However, in the case of QCD (similar to
the case of an antiferromagnet \cite{RajagopalWilczek,Hohenberg}), since the
dispersion relation of pions is given in terms of static quantities
only,
it turns out that $z$, as one would expect, can be derived from
static scaling laws only.

To determine $z$ we observe that the dispersion relation $\omega\sim
|\p|^z$ applies at scales $m_\sigma\ll |\p| \ll T$. At softer scales,
$|\p|\ll m_\sigma$, the dispersion relation is still linear,
$\omega=u |\p|$. Requiring that the two expressions for $\omega$
match at $|\p|\sim m_\sigma$ we find
\begin{equation}
m_\sigma^z \sim u\, m_\sigma \sim m_\sigma^{d/2},
\end{equation}
where to obtain the last scaling relation we used Eq.\ (\ref{umsigma}).
Therefore, at the QCD chiral phase transition the dynamical
critical exponent is given by 
\begin{equation}
z={d\over2}\,
\end{equation}
(see also an alternative derivation in Refs.\ 
\cite{RajagopalWilczek,Hohenberg}).\footnote{
This result agrees with 
\cite{RajagopalWilczek,Hohenberg}, but disagrees with \cite{Boyanovsky}.}
% where $z\stackrel{?}{=}1+\O(\epsilon)$ is obtained using expansion in
% $\epsilon=4-d$.  It is clear that if $z=d/2=2-\epsilon/2$, an attempt
% to expand in powers of $\epsilon$ assuming $z=1$ at $\epsilon=0$ must
% break down. In the calculation of the 1PI 2-point 
% function $\Sigma(p,\omega)$ in
% \cite{Boyanovsky} this is manifested by the correction to the free
% value $\omega^2-p^2$ of the order of $(g^2/\epsilon^2)\omega^2$. 
% Since the fixed
% point of the coupling $g$ is at $g=\O(\epsilon)$, the correction is of
% the same order as the leading term $\omega^2$. This indicates the
% breakdown of the $\epsilon$-expansion in \cite{Boyanovsky}.}

A similar argument can be applied to determine the scaling of the
diffusion coefficient $D'=D+\kappa_2$, characterizing the pion damping
in the chiral limit [Eq.\ (\ref{disp-im})]. The small momentum expansion of
$\omega$ is $u|\p|-{i\over2}D'\p^2 + \ldots$\ . The scaling hypothesis
dictates that all terms in this expansion become of the same order in
magnitude when $|\p|\sim m_\sigma$ (otherwise, another scale, e.g.,
$u/D'$, appears in addition to the inverse correlation length $m_\sigma$). This
requires
\begin{equation}
D' \sim m_\sigma^{(d-4)/2}\sim t^{\nu(d-4)/2}.
\end{equation}
Thus $D'$ diverges near the chiral phase transition in QCD ($d=3$).


\begin{thebibliography}{100}

\bibitem{dilepton}
%\bibitem{Li:1995qm}
G.~Q.~Li, C.~M.~Ko, and G.~E.~Brown,
%``Enhancement of low mass dileptons in heavy ion collisions,''
Phys.\ Rev.\ Lett.\  {\bf 75}, 4007 (1995)
[nucl-th/9504025];
%%CITATION = NUCL-TH 9504025;%%
%\bibitem{Li:1996db}
%G.~Q.~Li, C.~M.~Ko and G.~E.~Brown,
%``Effects of in-medium vector meson masses on low-mass dileptons from SPS  heavy-ion collisions,''
Nucl.\ Phys.\ {\bf A606}, 568 (1996)
[nucl-th/9608040];
%%CITATION = NUCL-TH 9608040;%%
%\bibitem{Cassing:1995zv}
W.~Cassing, W.~Ehehalt, and C.~M.~Ko,
%``Dilepton production at SPS energies,''
Phys.\ Lett.\ B {\bf 363}, 35 (1995)
[hep-ph/9508233].
%%CITATION = HEP-PH 9508233;%%

\bibitem{lattice-dilepton}
%\bibitem{Karsch:2001uw}
F.~Karsch, E.~Laermann, P.~Petreczky, S.~Stickan, and I.~Wetzorke,
%``A lattice calculation of thermal dilepton rates,''
Phys.\ Lett.\ B {\bf 530}, 147 (2002)
[hep-lat/0110208].
%%CITATION = HEP-LAT 0110208;%%

\bibitem{BijlFeynman} A.~Bijl, Physica (Amsterdam) {\bf 8}, 655 (1940);
R.~P.~Feynman, in {\em Progress in Low Temperature Physics},
edited by C.~J.~Gorter (North-Holland, Amsterdam, 1955), Vol. 1; see also
R.~P.~Feynman, {\em Statistical Mechanics} (Benjamin, Reading, MA, 1972)

\bibitem{HHPR69} B.~I.~Halperin and P.~C.~Hohenberg,
%``Hydrodynamic theory of spin waves''
Phys. Rev. {\bf 188}, 898 (1969).

%\cite{Pisarski:1984ms}
\bibitem{PisarskiWilczek}
R.~D.~Pisarski and F.~Wilczek,
%``Remarks On The Chiral Phase Transition In Chromodynamics,''
Phys.\ Rev.\ D {\bf 29}, 338 (1984).
%%CITATION = PHRVA,D29,338;%%

%\cite{Rajagopal:1993qz}
\bibitem{RajagopalWilczek}
K.~Rajagopal and F.~Wilczek,
%``Static and dynamic critical phenomena at a second order QCD 
%phase transition, '' 
Nucl.\ Phys.\ {\bf B399}, 395 (1993)
[hep-ph/9210253].
%%CITATION = HEP-PH 9210253;%%


%\cite{Hansen:1991yg}
\bibitem{Hansen}
F.~C.~Hansen and H.~Leutwyler,
%``Charge correlations and topological susceptibility in QCD,''
Nucl.\ Phys.\ {\bf B350}, 201 (1991).
%%CITATION = NUPHA,B350,201;%%


\bibitem{Smilga}
A.~V.~Smilga,
%``Physics of thermal QCD,''
Phys.\ Rep.\  {\bf 291}, 1 (1997)
[hep-ph/9612347].
%%CITATION = HEP-PH 9612347;%%

\bibitem{critical_pions}
D.~T.~Son and M.~A.~Stephanov,
%``Pion propagation near QCD critical point,''
Phys.\ Rev.\ Lett.\ {\bf 88}, 202302 (2002)
[hep-ph/0111100].
%%CITATION = HEP-PH 0111100;%%

\bibitem{mu2}
J.~B.~Kogut, M.~A.~Stephanov, and D.~Toublan,
%``On two-color QCD with baryon chemical potential,''
Phys.\ Lett.\ B {\bf 464}, 183 (1999)
[hep-ph/9906346];
%%CITATION = HEP-PH 9906346;%%
J.~B.~Kogut, M.~A.~Stephanov, D.~Toublan, J.~J.~Verbaarschot, 
and A.~Zhitnitsky,
%``QCD-like theories at finite baryon density,''
Nucl.\ Phys.\ {\bf B582}, 477 (2000)
[hep-ph/0001171].
%%CITATION = HEP-PH 0001171;%%

\bibitem{mui}
D.~T.~Son and M.~A.~Stephanov,
%``QCD at finite isospin density,''
Phys.\ Rev.\ Lett.\  {\bf 86}, 592 (2001)
[hep-ph/0005225];
%%CITATION = HEP-PH 0005225;%%
Yad.\ Fiz.\  {\bf 64}, 899 (2001)
[Phys.\ At.\ Nucl.\  {\bf 64}, 834 (2001)]
[hep-ph/0011365].
%%CITATION = HEP-PH 0011365;%%

\bibitem{gottlieb}
%\bibitem{Gottlieb:1987ac}
S.~Gottlieb, W.~Liu, D.~Toussaint, R.~L.~Renken, and R.~L.~Sugar,
%``The Quark Number Susceptibility Of High Temperature QCD,''
Phys.\ Rev.\ Lett.\  {\bf 59}, 2247 (1987);
%%CITATION = PRLTA,59,2247;%%
%\bibitem{Gottlieb:1997ae}
S.~Gottlieb, % {\it et al.},
U.~M.~Heller, A.~D.~Kennedy, S.~Kim,
J.~B.~Kogut, C.~Liu, R.~L.~Renken, D.~K.~Sinclair, R.~L.~Sugar, 
D.~Toussaint, and K.~C.~Wang,
%``Thermodynamics of lattice QCD with two light quark flavours on a  
%16**3 x 8 lattice. II,''
Phys.\ Rev.\ D {\bf 55}, 6852 (1997)
[hep-lat/9612020].
%%CITATION = HEP-LAT 9612020;%%

%\bibitem{Shuryak:1990ie}
\bibitem{Shuryak}
E.~V.~Shuryak,
%``Physics Of The Pion Liquid,''
Phys.\ Rev.\ D {\bf 42}, 1764 (1990).
%%CITATION = PHRVA,D42,1764;%%



%\cite{Goity:1989gs}
\bibitem{Goity}
J.~L.~Goity and H.~Leutwyler,
%``On The Mean Free Path Of Pions In Hot Matter,''
Phys.\ Lett.\ B {\bf 228}, 517 (1989).
%%CITATION = PHLTA,B228,517;%%

\bibitem{hydro_corr}
L.~P.~Kadanoff and P.~C.~Martin,
%``Hydrodynamic equations and correlation functions,''
Ann. Phys. (N.Y.) {\bf 24}, 419 (1963); D.~Forster, {\em Hydrodynamic
Fluctuations, Broken Symmetry, and Correlation Functions} (Benjamin,
Reading, MA, 1975).

\bibitem{Yaffe}
L.~G.~Yaffe (private communication).

\bibitem{Sonhydro}
%\bibitem{Son:2000pa}
D.~T.~Son,
%``Hydrodynamics of nuclear matter in the chiral limit,''
Phys.\ Rev.\ Lett.\  {\bf 84}, 3771 (2000)
[hep-ph/9912267].
%%CITATION = HEP-PH 9912267;%%

\bibitem{Hohenberg}
P.~C.~Hohenberg and B.~I.~Halperin,
%``Theory Of Dynamic Critical Phenomena,''
Rev.\ Mod.\ Phys.\  {\bf 49}, 435 (1977).
%%CITATION = RMPHA,49,435;%%

%\cite{Pisarski:1996mt}
\bibitem{Pisarski}
R.~D.~Pisarski and M.~Tytgat,
%``Propagation of Cool Pions,''
Phys.\ Rev.\ D {\bf 54}, 2989 (1996)
[hep-ph/9604404].
%%CITATION = HEP-PH 9604404;%%

\bibitem{Toublan}
%\bibitem{Toublan:1997rr}
D.~Toublan,
%``Pion dynamics at finite temperature,''
Phys.\ Rev.\ D {\bf 56}, 5629 (1997)
[hep-ph/9706273].
%%CITATION = HEP-PH 9706273;%%

\bibitem{Kapusta}
J.~I.~Kapusta, {\em Finite Temperature Field Theory} (Cambridge
University Press, Cambridge, U.K., 1989).

%\bibitem{Pines} 
%See, e.g., D.~Pines and P.~Nozi\`eres, {\em The Theory of Quantum
%Liquids} (Benjamin, New York, 1966).

\bibitem{Boyanovsky}
%\bibitem{Boyanovsky:2001pa}
D.~Boyanovsky and H.~J.~de Vega,
%``Dynamics near the critical point: The hot renormalization group in  
%quantum field theory,''
Phys.\ Rev.\ D {\bf 65}, 085038 (2002)
[hep-ph/0110012].
%%CITATION = HEP-PH 0110012;%%

\end{thebibliography}
\end{document}